\documentclass{WileyMSP-template}

\usepackage[utf8]{inputenc} 

\usepackage{amsmath}                 
\usepackage{amssymb}                 
\usepackage{mathrsfs}                
\usepackage{mathtools}               
\usepackage{dsfont}                  

\usepackage{microtype}

\usepackage{siunitx}    
\usepackage{suffix}     
\usepackage{physics}    

\usepackage[unicode]{hyperref} 
\usepackage[noabbrev,capitalise]{cleveref} 

\usepackage{geometry}

\newcommand{\PHDG}{^{\vphantom{\dagger}}}
\renewcommand{\vec}[1]{\mathrm{\bf{#1}}}

\begin{document}

\pagestyle{fancy}
\rhead{\includegraphics[width=2.5cm]{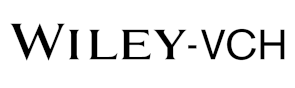}}

\title{Fermi-Surface Modeling of Light-Rare-Earth Hexaborides \\ with 2D-ACAR Spectroscopy}

\maketitle


\author{Josef Ketels*}
\author{Michael Leitner}
\author{Peter B\"oni}
\author{Christoph Hugenschmidt*}
\\
\author{Mikheil Sekania*}
\author{Alyn D. N. James}
\author{Jakob A. E. Bonart}
\author{Nico Unglert}
\author{Liviu Chioncel*}


\begin{affiliations}
J.~Ketels, P.~B\"oni\\
Physik-Department, Technische Universität München, James-Franck-Straße 1, 85748 Garching, Germany\\
Email Address: Josef.Ketels@frm2.tum.de

M.~Leitner, C.~Hugenschmidt\\
Physik-Department, Technische Universität München, James-Franck-Straße 1, 85748 Garching, Germany\\
Heinz Maier-Leibnitz Zentrum (MLZ), Technische Universit\"at M\"unchen, Lichtenbergstra{\ss}e 1, Garching, 85748, Germany\\
Email Address: Christoph.Hugenschmidt@frm2.tum.de

M.~Sekania\\
Theoretical Physics III, Center for Electronic Correlations and Magnetism, Institute of Physics, University of Augsburg, Universit\"atsstra{\ss}e 1, 86159 Augsburg, Germany\\
Institut f\"ur Physik, Martin-Luther Universit\"at Halle-Wittenberg, 06120 Halle/Saale, Germany\\
Andronikashvili Institute of Physics, Javakhishvili Tbilisi State University, Tamarashvili str. 6, 0177 Tbilisi, Georgia\\
Email Address: Mikheil.Sekania@physik.uni-halle.de

A.~D.~N.~James\\
H.~H.~Wills Physics Laboratory, University of Bristol, Tyndall Avenue, Bristol, BS8 1TL, United Kingdom

N.~Unglert, J.~A.~E.~Bonart, L.~Chioncel \\ 
Theoretical Physics III, Center for Electronic Correlations and Magnetism, Institute of Physics, University of Augsburg, Universit\"atsstra{\ss}e 1, 86159 Augsburg, Germany\\
Augsburg Center for Innovative Technologies, University of Augsburg, Universit\"atsstra{\ss}e 1, 86159 Augsburg, Germany\\
Email Address: Liviu.Chioncel@physik.uni-augsburg.de
\end{affiliations}



\begin{abstract}
Two dimensional angular correlation of the positron annihilation radiation  (2D-ACAR) spectra are measured for $\mathrm{LaB}_6$ along high symmetry directions and compared with first principle calculations based on density functional theory (DFT).
This allows the modeling of the Fermi surface in terms of ellipsoid electron pockets centered at $X$-points elongated along the $\Sigma$ axis (${\Gamma-M}$ direction). The obtained structure is in agreement with quantum oscillation measurements and previous band structure calculations.
For the isostructural topologically not-trivial $\mathrm{SmB}_6$ the similar ellipsoids are connected through necks that have significantly smaller radii in the case of $\mathrm{LaB}_6$.
A theoretical analysis of the 2D-ACAR spectra is also performed for $\mathrm{CeB}_6$ including the on-site repulsion $U$ correction to the local-density approximation (LDA+$U$) of the DFT.
The similarities of 2D-ACAR spectra and the Fermi-surface projections of these two compounds allow to infer that both $\mathrm{LaB}_6$ and $\mathrm{CeB}_6$ are topologically trivial correlated metals.
\end{abstract}


\justifying

\section{Introduction}
\label{sec:intro}

The rare-earth hexaborides ($R\mathrm{B_6}$, $R=\mathrm{La}$, $\mathrm{Ce}$, $\mathrm{Pr}$, $\mathrm{Nd}$, and $\mathrm{Sm}$) provide an interesting subject for experimental and theoretical studies because of a variety of features that can be ascribed to the particular role played by their $4f$-electrons~\cite{de.fe.97,rise.00}. 
The unfilled $4f$-states are a challenging problem for an accurate theoretical description of the electronic structure of rare-earth compounds. The $4f$ energy levels often overlap with the non-$4f$ broad bands, form narrow resonances, but nevertheless are frequently treated as core states. A proper description, however, requires inclusion of the $4f$-states in valence bands where these states are also subjected to a strong on-site Coulomb repulsion.
A simple theory that can capture this strong on-site Coulomb repulsion for $4f$ orbitals at the mean-field level is the LDA+$U$ approach~\cite{an.ar.97}.  
Another important ingredient in the physics of the rare-earth compounds is the presence of significant spin-orbit coupling, which fortunately can be included on an equal footing with the strong on-site Coulomb interactions (the Hubbard $U$). The recent theoretical models beyond the local approximations, such as $GW$~\cite{ar.gu.98}, or dynamic electronic correlations such as Dynamical Mean Field Theory (DMFT)~\cite{ko.vo.04,ko.sa.06}, however, are difficult to use in case of the multi-orbitals (like $4f$) and the strong spin-orbit couplings.     

In the $R\mathrm{B}_6$ series of compounds, $\mathrm{LaB}_6$ is considered as the reference non $f$-electron system, $\mathrm{CeB}_6$ is a Kondo lattice heavy fermion system, while $\mathrm{PrB}_6$ and $\mathrm{NdB}_6$ are local-moment magnetically ordered metals.
Among these compounds $\mathrm{SmB}_6$ is considered to be a ``topological Kondo insulator''~\cite{dz.su.10} due to the narrow hybridization band gap that opens at the Fermi level.
In general, some of the $4f$-bands of the rare-earth elements are located around the Fermi level ($E_{\mathrm{F}}$) and are crossed by the more dispersive $5d$ band.
The finite spin-orbit coupling splits these crossings and opens a gap.
The band-character inversion due to the spin-orbit coupling between the rare-earth $4f$/$5d$ states situated around $E_{\mathrm{F}}$ signals the presence of a topologically non-trivial phase.
This type of behavior was observed in DFT calculations for $\mathrm{SmB}_6$~\cite{ch.ta.15} and further searched in other rare-earth hexaborides~\cite{sh.ho.20}.
The effects due to a finite Hubbard interaction, $U$, were also investigated for $\mathrm{Sm}$-$4f$ and it was found that the topological features in $\mathrm{SmB}_6$ are insensitive to the values of the on-site Coulomb interaction.
The latter, however, plays a crucial role in the other isostructural hexaborides~\cite{sh.ho.20}.

In the present study we investigate the electronic properties of two members of the $R\mathrm{B}_6$ family, namely $\mathrm{LaB}_6$ and $\mathrm{CeB}_6$, with two dimensional angular correlation of annihilation radiation (2D-ACAR) and first principle Density Functional Theory (DFT)~\cite{ho.ko.64,kohn.99,jo.gu.89,jone.15}.
We analyze the radial anisotropies of the obtained data in the extended momentum space and the corresponding back-folded spectra in the crystal-momentum space in order to identify the different Fermi surface features.
We also compare the experimentally measured and computed 2D-ACAR spectra for $\mathrm{LaB}_6$.
The crystal structures of the $R\mathrm{B}_6$ belong to the simple cubic $\mathrm{CsCl}$ structure type ($Pm\overline{3}m$), with lattice constant ${a=\SI{4.1569}{\angstrom}}$ and ${a=\SI{4.1391}{\angstrom}}$, for $\mathrm{LaB}_6$ and $\mathrm{CeB}_6$, respectively. The rare-earth atoms occupy the corner of the unit cell, corresponding to the ${1a\,(0, 0, 0)}$ Wyckoff site, while the B atoms are located at the octahedral sites in the body-centered position at the ${6f\,(0.5, 0.5, z)}$ Wyckoff sites, where $z$ is \num{0.2011} for $\mathrm{CeB}_6$~\cite{Tanaka2002} and \num{0.1996} for $\mathrm{LaB}_6$ \cite{CHEN2004}.

Although the $f$-shell of $\mathrm{La}$ is empty while $\mathrm{Ce}$ has a single $f$-electron and $\mathrm{CeB}_6$ exhibits a far richer phase diagram than $\mathrm{LaB}_6$, undergoing several different magnetic phase transitions, de Haas-Van Alphen (dHvA) measurements~\cite{jo.ru.87} yielded similar Fermi surface (FS) features for both $\mathrm{CeB}_6$ and $\mathrm{LaB}_6$.
The FS of $\mathrm{CeB}_6$ consists of large ``ellipsoid'' pockets around the $X$ points of the Brillouin zone (BZ), whereas the states around the zone center ($\Gamma$-point) are shifted away by band-renormalization effects leading to a hole pocket~\cite{ne.al.15}.
The first-principles calculations agree with the experimental results around the $X$-points but fail to capture the strongly renormalized electronic states around the $\Gamma$-point~\cite{ne.al.15}.
The dHvA measurements~\cite{jo.ru.87} at temperatures ${T<T_K}$
can be brought into accordance with a model in which $f$-electrons do not contribute to the Fermi volume.
Above the Kondo temperature ${T>T_K}$, in the paramagnetic phase, dHvA measurements and 2D-ACAR results were also in agreement in the case of $\mathrm{CeB}_6$~\cite{jo.ru.87}.
Results of the 2D-ACAR experiments on $\mathrm{LaB}_6$ were reported earlier in the literature~\cite{bi.al.96,bi.fr.97}, where 
the analysis in terms of the back-folded momentum densities were performed.
Three-dimensional reconstruction of the Fermi surface using 2D-ACAR projections were also reported~\cite{bi.mo.01}, and
contrary to the electronic structure calculations the strong hybridization between $\mathrm{B}$-$p$ and $\mathrm{La}$-$d$ states ($f$-states are unoccupied) that produced additional FS sheets, were not detected in the experiment.  

In our work we analyze the momentum densities of $\mathrm{LaB}_6$ (experiment/theory) both in the extended $p$ and reduced $k$ spaces. Possible similarities between the theoretical spectra with the isostructural $\mathrm{CeB}_6$ are also discussed. For the $\mathrm{CeB}_6$ compound, in the absence of recent experimental measurements, we theoretically study the effects of the different $U$ values of the Hubbard interaction and diverse double counting schemes of the LDA+$U$ method.
Based on our 2D-ACAR results, we conclude that both $\mathrm{LaB}_6$ and $\mathrm{CeB}_6$ are topologically trivial but correlated metals.
Therefore, we believe that the future studies with DMFT which takes into account the full local correlations starting from itinerant (valence) $4f$-states will provide a suitable description of the electronic structure of the rare-earth hexaborides.     

\section{Methods}

\subsection{2D-ACAR}
\label{subsec:2D-ACAR}
2D-ACAR is a powerful tool to investigate the bulk electronic structure~\cite{Weber2015,Ceeh2016}. It is based on the annihilation of positrons with electrons of a sample leading to the emission of two $\gamma$-quanta in nearly anti-parallel directions. The small angular deviation from collinearity is caused by the transverse component of the electron's momentum. Hence, the coincident measurement of the annihilation quanta for many annihilation events yields a projection  of the so called two photon momentum density (TPMD) $\rho^{2\gamma}(\vec{p})$.
This is usually computed as the Fourier transform of the product of positron wave function 
$\Psi^+(\vec{x})$ and electron wave function $\Psi^-({\vec{x}})$:
\begin{equation}
	\rho^{2\gamma}({\vec{p}})\propto \sum_{j,\vec{k}} n_j({\vec{k}}) \left| \int \mathrm{d} \vec{x} \, e^{-i {\vec{x p}}} \, \Psi^+({\vec{x}}) \Psi^-({\vec{x}}) \, \sqrt{\gamma({\vec{x}})} \right| ^2
    \label{eq:rho2gamma}
\end{equation}
The sum runs over all $\vec{k}$ states in all bands, $j$, with the occupation $n_j(\vec{k})$. The so-called ``enhancement factor'' $\gamma({\vec{x}})$~\cite{Jarlborg1987}, takes into account the electron positron correlation. The 2D-ACAR spectrum $\rho\PHDG_{\mathrm{ACAR}}(p_x,p_y)$, the quantity which is actually accessible by an experiment, is a 2D projection of the 3D momentum-density distribution $\rho^{2\gamma}(\vec{p})$ along a chosen ($p_z$) axis
\begin{equation}
\label{eq:2dacar}
\rho\PHDG_{\mathrm{ACAR}}(p_x,p_y) =  \int \rho^{2\gamma}({\vec{p}}) \mathrm{d} p_z\,.
\end{equation}

The positron annihilation probes all electrons in the system. Filled bands, especially bands of core electrons, give a nearly isotropic distribution which is superimposed by an anisotropic contribution mainly produced by the electrons near the Fermi surface. This anisotropic  $\rho\PHDG_{\mathrm{aniso}}(p_x,p_y)$  contribution is therefore one of the most interesting feature of an ACAR spectrum $\rho\PHDG_{\mathrm{ACAR}}(p_x,p_y)$. It can be calculated by subtracting the isotropic contribution ${\overline{\rho}\PHDG_{\mathrm{ACAR}}(p_x,p_y)}$:
\begin{equation}
\label{eq:A}
\rho\PHDG_{\mathrm{aniso}}(p_x,p_y) = \rho\PHDG_{\mathrm{ACAR}}(p_x,p_y)  -\overline{\rho}\PHDG_{\mathrm{ACAR}}(p_x,p_y)\,,
\end{equation}
where the isotropic contribution is the radial average ${\overline{\rho}\PHDG_{\mathrm{ACAR}}(p_x,p_y) \equiv \overline{\rho}\PHDG_{\mathrm{ACAR}}(\sqrt{p_x^2+p_y^2})}$ constructed from the original spectrum $\rho\PHDG_{\mathrm{ACAR}}(p_x,p_y)$ by averaging over all data points in equidistant intervals ${[p_r,p_r+\Delta p_r)}$ from the center.

The discontinuities in $\rho^{2\gamma}(\vec{p})$ correspond to the density jumps in the Fermi distribution and are used to identify the Fermi surface sheets. These discontinuities occurring at ${\vec{p}=\vec{k}+\vec{G}}$, where $\vec{k}$ is the wave vector in the first BZ and $\vec{G}$ is a reciprocal lattice vector, can be folded back from the extended zone scheme $p$-space to the reduced zone scheme $k$-space by applying the so called Lock-Crisp-West (LCW) theorem~\cite{Lock1973}. 

\subsection{Experiment Details}

The ACAR experiments were performed at the 2D-ACAR spectrometer at the {\em Technische Universit\"at München}. The detector-detector distance is \SI{17.5}{\meter} with the sample positioned exactly in the middle. This leads to an angular resolution of ${\sigma_h = \SI{0.538(1)}{mrad}}$ and  ${\sigma_v = \SI{0.655(1)}{mrad}}$ in horizontal and vertical direction, respectively~\cite{Leitner2012}. All measurements were conducted at a temperature of \SI{15}{\kelvin} to reduce the contribution of the thermal motion of the positron. The positrons were guided onto the sample by a magnetic field of \SI{1.2}{\tesla} at the sample position. For full details on the experimental setup we refer to ref~\cite{Ceeh2013}. 

The $\mathrm{LaB}_6$ sample used in the 2D-ACAR experiment was a cuboid-shaped high-quality single crystal of ${10\times5\times2}$~\si{\milli\meter\cubed} size. It was grown by vertical crucible-free inductive floating zone melting in argon gas atmosphere. The grown crystal was characterized by Laue back-scattering, optical spectral analysis, X-ray diffraction and density measurements. The surfaces of the plate were polished with diamond powder. 

We recorded 2D-ACAR projections along the high symmetry $[110]$, $[001]$ and $[111]$ directions within the $(1\bar{1}0)$ plane. For each spectrum more than \SI{9e7}{counts} were collected. To correct for the varying efficiency and limited aperture of the Anger camera detectors, the raw data were divided by the so-called momentum sampling function (MSF). To calculate the MSF the single events on both detectors were collected in parallel to the coincident events throughout the experiment. The convolution of those single spectra gives the MSF. The ACAR spectra were symmetrized according to the crystal point-symmetry to further enhance the statistics.

\subsection{Electronic Structure Calculation and Enhancement Models}

Electronic structure calculations were performed using the ELK code~\cite{elk}, which is an all-electron full-potential linearised augmented plane-wave (LAPW) code for determining the properties of crystalline solids.
The valence electron configuration for the rare-earth 
atoms ($\mathrm{La/Ce}$) is ${6s6p5d4f}$, while the valence electrons of B are located in the $2s$ and $2p$ orbitals. The self-consistent calculations 
were performed using the Local Density Approximation (LDA) exchange-correlation functional parameterized by Perdew and Wang~\cite{pe.wa.92}. For the LDA+$U$ the rationally invariant formulation of the Coulomb interaction between the valence electrons was used~\cite{an.ar.97} with different double counting schemes~\cite{pe.ma.03}.
Further details of the LDA+$U$ calculations can be found in \cref{subsec:LDA+U}.
In the self-consistent calculations the $\vec{k}$-summations are performed using the tetrahedron method, with a $20\times20\times20$ $k$-mesh in the irreducible part of the Brillouin zone and convergence has been achieved in total energy with an accuracy of better than $10^{-5}$~Ry. For the detailed Fermi surface calculations a significantly larger, $80\times80\times80$, $k$-mesh was employed.

The DFT can be generalized to electron-positron systems by including the positron density, in the form of the two-component DFT~\cite{bo.ni.86,pu.ni.94}. 
In this formalism the positron is considered to be thermalized and described by a state with ${{\vec{p}}_p =0}$ with $s$-type symmetry at the bottom of the positronic band.
The photons resulting from the electron-positron annihilation carry the momentum of the annihilated pair up to a reciprocal lattice vector, reflecting the fact that the annihilation takes place in a crystal.
Hence, an electron with a transverse wave vector ${\vec{k}}$ contributes to spin-resolved two-photon momentum density, $\rho^{2\gamma}(\textbf{p})$ not only at ${{\vec{p}} = {\vec{k}}}$ (normal process) but also at ${{\vec{p}} = {\vec{k}} + {\vec{G}}}$, with ${\vec{G}}$ a vector of the reciprocal lattice (Umklapp process).
In the LDA(+$U$) framework the electron-positron momentum density $\rho({\vec{p}})$ is computed directly with the spin-resolved versions of \cref{eq:rho2gamma,eq:2dacar}. 
We calculate the 2D-ACAR spectra according to the method described in ref~\cite{er.bi.14}. 
These TPMD were calculated to a maximum momentum of $6$~a.u. and the
electron positron correlations are taken into account within the $\gamma(\vec{x})$ term, the so-called enhancement factor.
Within the Independent Particle Model (IPM) ${\gamma(\vec{x})=1}$, while models of enhancement beyond IPM consider momentum and energy dependencies 
and a separate treatment of $f$, $d$ states in comparison with the $s$, $p$ states. These have been quantified via many different approximations~\cite{la.ha.10}. In the present studies we employ the so called Drummond parametrization~\cite{dr.lo.10,dr.lo.11}.

\begin{figure}[!tbp]
    \includegraphics[width=\textwidth]{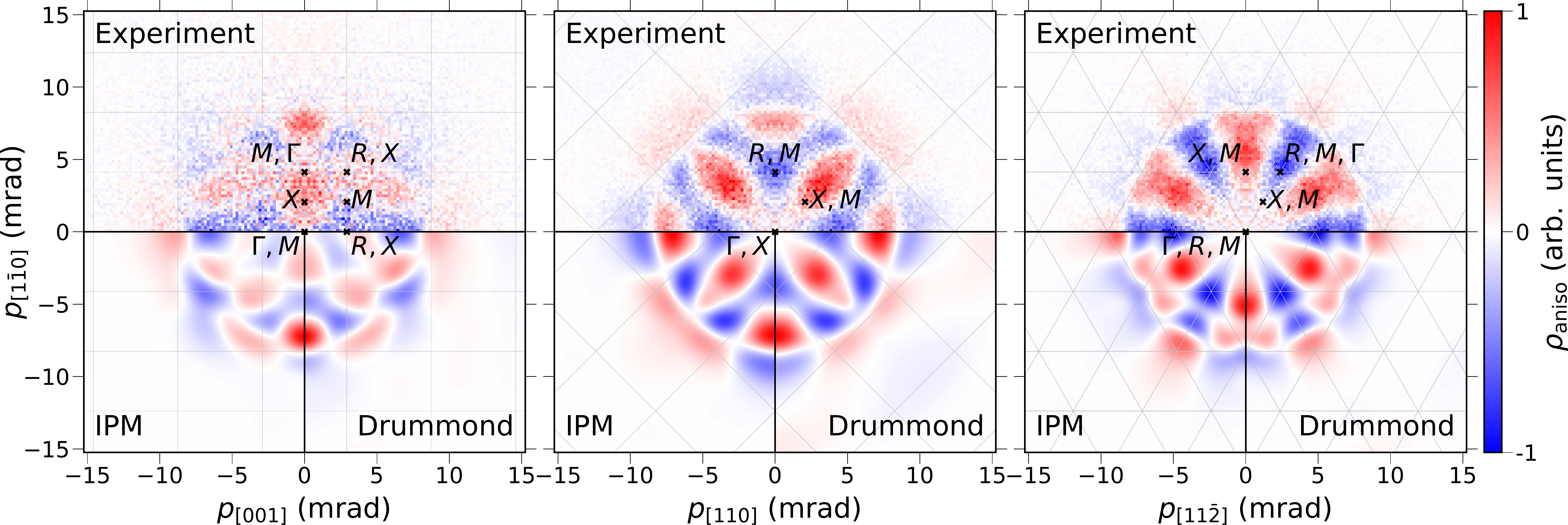}
    \caption{Radial anisotropy of measured and symmetrized $\mathrm{LaB}_6$  ACAR spectra (top) and theoretical spectra calculated in the IPM (bottom left) and with the Drummond enhancement (bottom right). The spectra from first-principle calculations are convolved with a Gaussian accounting for the experimental resolution of ${\SI{0.619}{\milli\radian}}$.
    The borders of the projected first BZ (grey lines) in a repeated zone scheme are shown in all plots.
    From left to right $(110)$, $(001)$, and $(111)$ high symmetry projections.}
    \label{fig:RA-acar-exp-vs-theory}
\end{figure}

\section{Results and Discussion}

\subsection{$\mathrm{LaB}_6$: Experiment and Theory}

To compare experimental and theoretical data on $\mathrm{LaB}_6$ in the $p$-space we look at the radial anisotropy  $\rho\PHDG_{\mathrm{aniso}}(p_x,p_y)$, which emphasizes the spectral contribution from electrons near the Fermi level. \cref{fig:RA-acar-exp-vs-theory} shows the radial anisotropy for all three measured projections (from left to right $(110)$, $(001)$ and $(111)$) in the upper half and the corresponding theoretical calculations in the lower half of the plots.
The calculated results are for the momentum distributions with IPM (left) and Drummond enhancement (right).
To account for the limited experimental resolution the first-principle spectra are convolved with a Gaussian of width ${\sigma\PHDG_{\mathrm{exp}}=\SI{0.619}{\milli\radian}}$. 

Overall, we have a very good agreement between theory and experiment. Dominant features seen in the theoretical calculations are also visible in the experimental data.
Especially the four-fold and sixfold symmetric projections are well reproduced.
Some minor differences can be seen in the $(110)$ projection. 
Particularly, the weight distribution in the region between $-5$ to $\SI{5}{\milli\radian}$ on both axes shows a significant departure between the theory and experiment.
However, the high intensity positive signal along ${[1\bar{1}0]}$ direction at about $\SI{8}{\milli\radian}$ is well reproduced. 
We attribute the better agreement between the theory and the experiment for the $(001)$ and $(111)$ projections to the higher point symmetry of these directions. 
In all experimental projections the anisotropy profiles are slightly stretched out within the plane.
The similar behaviour is also observed when the Drummond enhancement is included in the computation as compared to the IPM results.
However, the radial expansion of the profile is not sufficient to match the observed ``stretching'' in the experimental measurement.
The differences between the IPM and Drummond model are marginal, though.

\begin{figure}[tbp]
    \includegraphics[width=\textwidth]{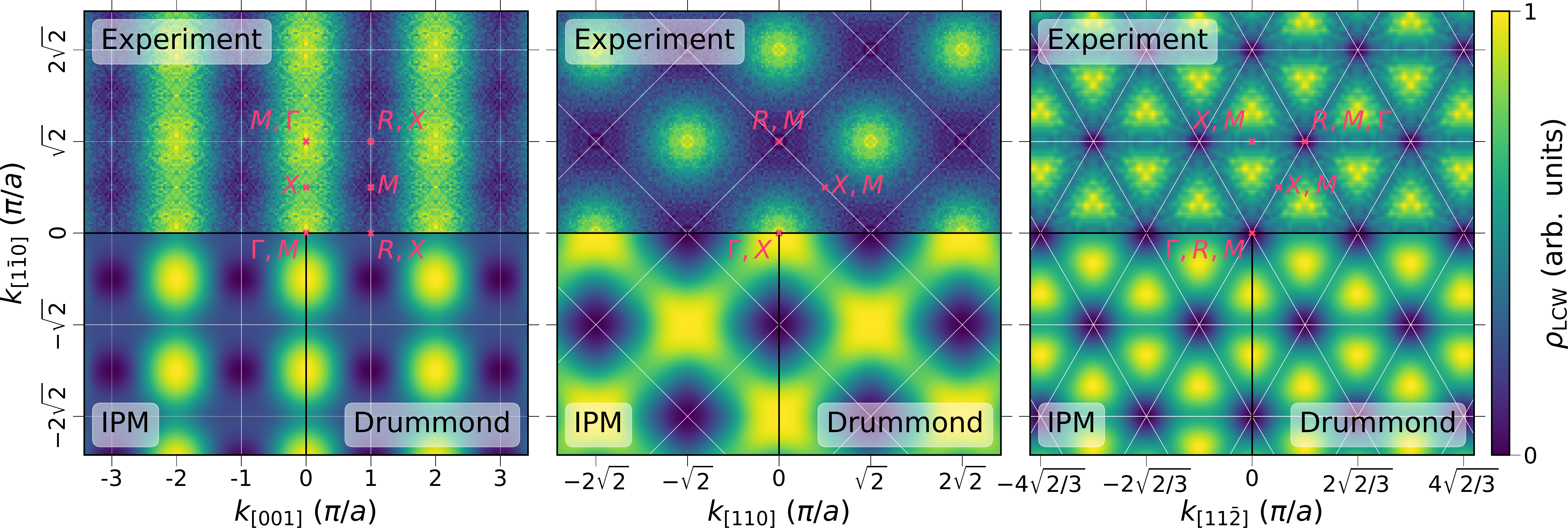}
    \caption{$\mathrm{LaB}_6$ LCW back-folded $k$-space density of the ACAR measurements (top) and the theoretical calculations in the IPM (bottom left) and the Drummond model (bottom right). The spectra from first-principle calculations are convolved with a Gaussian accounting for the experimental resolution of ${\SI{0.619}{\milli\radian}}$. 
    The borders of the projected first BZ (white lines) in a repeated zone scheme are shown in all plots.
    From left to right $(110)$, $(001)$, and $(111)$ high symmetry projections.}
    \label{fig:lcw-acar-exp-vs-theo}
\end{figure}

\cref{fig:lcw-acar-exp-vs-theo} shows the LCW back-folded data, experiments as well as theory, and is organized according to the structure of \cref{fig:RA-acar-exp-vs-theory}.
The top part of each plot shows the $k$-space experimental 2D-ACAR spectrum, while the bottom part is split between the calculated results using IPM (left) and Drummond enhancement (right), broadened with the experimental resolution. 

\begin{figure}[tbp]
    \centering
    \includegraphics[width=\textwidth]{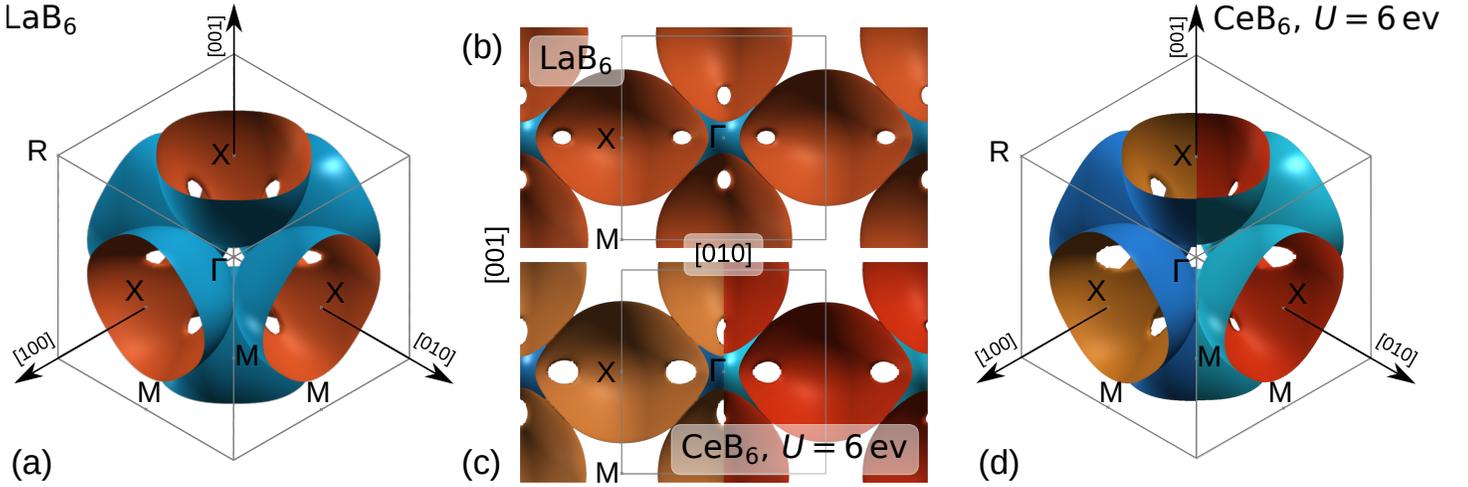} 
    \caption{Fermi surface in the first BZ (a),(d) and ${\Gamma-X-M}$ cross-section (${[010]-[001]}$ cut) (b),(c) from first-principle calculations for $\mathrm{LaB}_6$ (a),(b) and for $\mathrm{CeB}_6$ (c),(d), respectively.
    FS sheets, for two different spin projections, are depicted in the left and the right half of the plots (c) and (d), respectively. 
    The borders of the first BZ (grey lines) and high-symmetry points ($\Gamma$, $X$, $M$, and $R$) are shown as well.
    }
    \label{fig:fermi-surf}
\end{figure}

The measured and calculated LCW back-folded spectra show an overall good agreement.
The $(001)$ and $(111)$ projections show the expected features of the Fermi surface topology (see \cref{fig:fermi-surf}).
In the $(111)$ projection, the sixfold rose structure around the $\Gamma$-point becomes apparent.
As expected from the calculated Fermi surface, there is no density at the $\Gamma$- and $R$-point in the BZ.
Note that one can not see the hole-space around the $\Gamma$-point in the $(001)$ projection, as they are covered by the  ``ellipsoids'' along the $[001]$ direction.
While the theoretical calculations explicitly reproduce the ``ellipsoids'' structure in $(110)$ projection, the experimental $(110)$ projection shows stripes along $[1\bar{1}0]$ direction. The latter can be also attributed to the observed substantial discrepancies between the theory and experiment in the radial anisotropy (cf. \cref{fig:RA-acar-exp-vs-theory}).

The Fermi surface of $\mathrm{LaB}_6$ is shown in \cref{fig:fermi-surf} in the simple cubic BZ where the high symmetry points are ${\Gamma\equiv[0,0,0]}$, ${X\equiv[1/2,0,0]}$, ${M\equiv[1/2,1/2,0]}$ and ${R\equiv[1/2,1/2,1/2]}$.
The Fermi surface of $\mathrm{LaB}_6$ consists of a  set of equivalent ``ellipsoids'' centered at the $X$-points and connected by necks which intersect along $\Sigma$ ($\Gamma-M$) direction.
In the panels (b) and (c) in \cref{fig:fermi-surf} we show a cross-section in the ${\Gamma-X-M}$-plane (${k_{[100]}=0}$ plane).
The DFT~(LDA) Fermi surface calculation for $\mathrm{SmB}_6$ with a downward shift of the Fermi energy is presented in ref~\cite{ta.hs.15} and has a FS similar to that of $\mathrm{LaB}_6$ as shown in \cref{fig:fermi-surf}). 
The angular variations of the dHvA frequencies and the disappearance of oscillations in some angular regions have been recently discussed for both $\mathrm{SmB}_6$ and $\mathrm{LaB}_6$~\cite{ta.hs.15}.
These quantum oscillation frequencies identify the so-called $\alpha,\gamma, \varepsilon$-branches associated to FS regions visible in the ${\Gamma-X-M}$-plane, see \cref{fig:fermi-surf}.
The $\alpha$-branches were associated with the ellipsoids centered at the $X$-point. 
As one can see the nearest-neighbor ellipsoid FS pieces touch along the ${\Gamma-M}$ line and connect through the small distorted circular shape neck building up a multiply connected FS. Both the $\gamma$ and $\varepsilon$-orbits are hole-like orbits and are centered around $M$-point and $\Gamma$-point, respectively. 
According to the dHvA frequencies the angular region for the hole-like $\gamma$- and $\varepsilon$-orbits is significantly larger in $\mathrm{SmB}_6$ than in $\mathrm{LaB}_6$ which may be the cause of the different physical properties of these two compounds. 
From our calculations we can compare linear dimensions of the Fermi surface features to both previous computations and the present experimental data. The corresponding values are $0.848$, $0.668$, and $0.624$ for the ${X-\Gamma}$, ${X-M}$, and ${X-R}$ directions, respectively, as a fraction of the BZ size. Our results agree with the previous results reported in the literature~\cite{bi.mo.01}. 

To determine the FS linear dimension from the experimental data~\cite{bi.mo.01,le.we.16}, taking into account a finite experimental resolution,
we model the electron-momentum density with prolate ellipsoids with equatorial radius $R_{\mathrm{eq}}$ and polar radius $R_{\mathrm{pl}}$, centered at the $X$ points in reciprocal space and pointing towards the neighbouring $\Gamma$ points
\begin{equation}
 F(\vec{k};R_{\mathrm{eq}},R_{\mathrm{pl}}) = 
 \sum\limits_{\vec{G}}
 \sum\limits_{\vec{b}_i}
 \Theta\left(
         1
         - \frac{|(\vec{k} - \vec{G})\times{\vec{b}_i}|^2}
                {R_{\mathrm{eq}}^2|\vec{b}_i|^2}
         - \frac{|(\vec{k} - \vec{G} - {\vec{b}_i}/2) \cdot {\vec{b}_i}|^2}
                {R_{\mathrm{pl}}^2|\vec{b}_i|^2}
      \right)\,.
\end{equation}
Here, $\Theta(\cdot)$ is the Heaviside step function and summation is over reciprocal lattice vectors $\vec{G}$ and reciprocal primitive vectors ${\vec{b}_i \in \{\vec{b}_{[100]},\vec{b}_{[010]},\vec{b}_{[001]}\}}$.
Adding a flat background two-photon momentum density due to fully occupied bands described by a third parameter $F_\mathrm{backg}$, projecting along the given direction and convolving with the anisotropic Gaussian kernel that models the experimental resolution, we fit the resulting simulated projections to the correspondingly normalized LCW back-folded data for all measured projection directions by varying the three free parameters. We obtain ${2R_\mathrm{eq}=0.662(2)}$ and ${2R_\mathrm{pl}=0.952(2)}$ in units of the reciprocal lattice constant,
where the reported uncertainties are the statistical precision due to counting noise. We see that the equatorial diameter $2R_\mathrm{eq}$ agrees well with the calculated ${X-M}$ length, while the length along the ${X-\Gamma}$ direction is overestimated.
We think that this is due to positron wave-function and/or electron-positron correlation effects, which are neglected in the LCW theorem~\cite{Lock1973}: While the equatorial diameter is directly accessible in particular in the (001) projection, where the contrast between the filled ellipsoids and the background density is large, there is no projection direction that would afford an unobstructed view on the polar termination of the ellipsoids. Thus, a larger band weight of the conduction and/or core bands towards $\Gamma$ would necessarily result in an overestimated $R_\text{pl}$. Also, with the current experimental resolution we cannot detect possible deviations of the Fermi surface from the shape of a rotational ellipsoid in the vicinity of the poles, or whether the ${X-M-R}$ cross-section of the FS shows four-fold deviations from circular symmetry, as shown by the calculations (see \cref{fig:fermi-surf}).

\subsection{$\mathrm{CeB}_6$: Theory}
\label{subsec:LDA+U}

Among the most frequently used techniques to include short-range Coulomb interactions between the electrons in the framework of DFT are the self-interaction-corrected local-spin-density approximation (SIC-LSDA)~\cite{pe.zu.81} and LDA+$U$~\cite{an.ar.97}. 
The LDA+$U$ method is particularly useful as it is computationally less demanding and hence can be easily used for systems with unit cells consisting of a considerable number of atoms. 
Furthermore, its flexibility allows a systematic study of effects of the on-site repulsion $U$, on the ground state properties such as equilibrium lattice parameter, magnetic moment, and --- in the current work --- 2D-ACAR spectra. 
The orbital dependent LDA+$U$ functional used in the present work is given by~\cite{an.ar.97}:
\begin{equation}
    E_{\mathrm{LDA}+U}=E_{\mathrm{LDA}}+\frac{U-J}{2}\sum_{\sigma}[\tr \rho^\sigma - \tr(\rho^\sigma \rho^\sigma)]
\end{equation}
Here, $\rho^\sigma$ (not to be confused with momentum densities) is the density matrix for the $f$-states and $U$ and $J$ are the local Coulomb and exchange Hund's parameters. 
The lack of a diagrammatic expansion of the DFT total energy makes it difficult to model the effect of the local Coulomb interactions beyond the effects already captured by the exchange-correlation functionals. 
To avoid the double counting of such effects several schemes have been proposed for different kind of materials.
One of them is the mean-field approximation to the Hubbard correction, the so-called ``fully-localized'' limit (FLL).
In this scheme each localized (e.g., atomic) orbital is either full or completely empty.
This formulation of the double-counting term mimics an expansion of the electronic energy around the strongly localized limit and thus tends to work quite well for strongly correlated materials with localized orbitals. 
For other systems, such as metals or ``weakly correlated'' materials, the excessive stabilization of occupied states due to the ``+$U$'' potential may lead to unphysical results, for example the enhancement of the Stoner factor~\cite{pe.ma.03}.
A different double-counting scheme, the ``around mean-field'' (AMF), was introduced in ref~\cite{cz.sa.94} and further developed in ref~\cite{pe.ma.03}.
In the present study the AMF double counting scheme failed to produce the correct position of the $4f$ bands, therefore FLL double-counting was employed in all presented results.

\begin{figure}[tbp]
    \includegraphics[width=\textwidth]{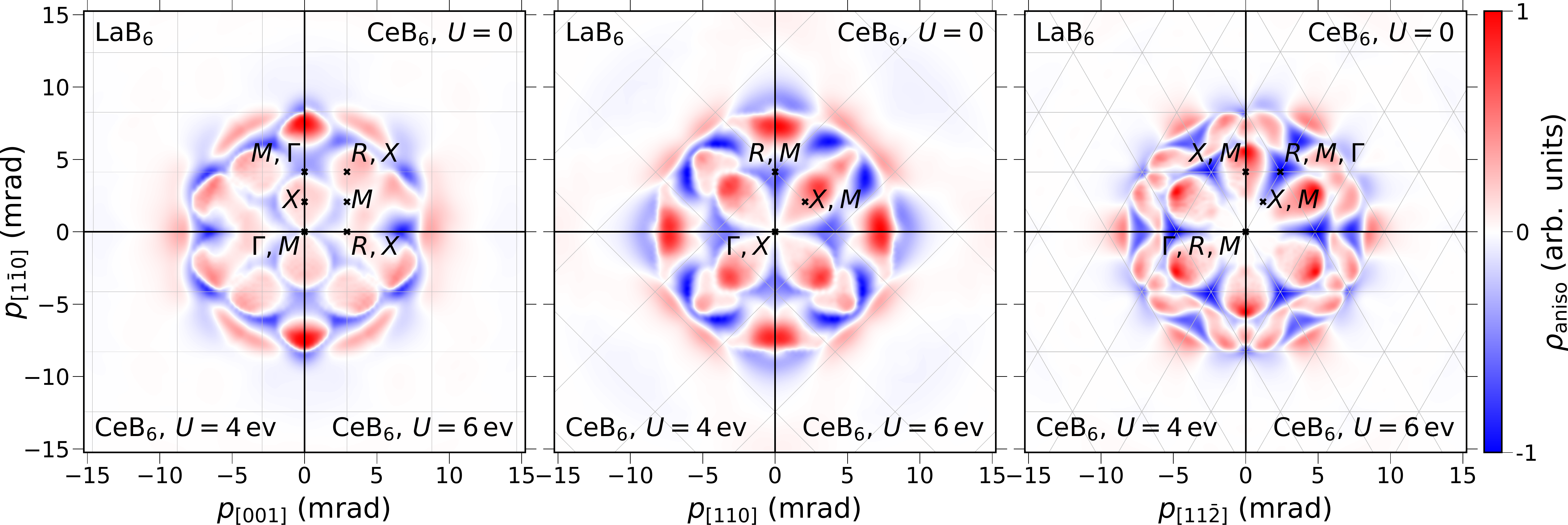}
    \caption{Unconvolved radial anisotropy of the calculated ACAR spectra for $\mathrm{LaB}_6$ (upper left) and $\mathrm{CeB}_6$ for ${U=0}$ (upper right), ${U=4}$~eV (bottom left), and ${U=6}$~eV (bottom right). Results correspond to LDA ($\mathrm{LaB}_6$ and $\mathrm{CeB}_6$ ${U=0}$) and LDA+$U$ (the rest) calculations with the Drummond parametrization of the electron-positron enhancement factor.
    The borders of the projected first BZ (grey lines) in the repeated zone scheme are shown in all plots.
    From left to right $(110)$, $(001)$, and $(111)$ high symmetry projections.}
    \label{fig:RA-CeB6-theo}
\end{figure}

The DFT band structure of $\mathrm{CeB}_6$ produces the manifold of $\mathrm{Ce}$-$4f$ orbitals in the close vicinity of the Fermi level.
According to the experimental study of Neupane~et~al.~\cite{ne.al.15} they have to be located at around $2$~eV in the conduction band.
Therefore, for the FLL limit a reasonable parameter for the effective Coulomb interaction would be $U=4$~eV.
The band structure and the density of states (not shown) in the ferromagnetic states are characterized by dispersive $5d$- and flat $4f$-bands along the ${M-X-M}$ and ${X-\Gamma-X}$ directions. The flat bands are purely build from $\mathrm{Ce}$-$4f$ states.
The dispersive $5d$ band around the $X$-point are situated $2$~eV below $E_{\mathrm{F}}$ and touches the $\mathrm{B}-2p$ states at the $\Gamma$ point. The position of these bands agrees fairly well with experiments~\cite{ne.al.15,ko.he.16}.

\begin{figure}[tbp]
    \includegraphics[width=\textwidth]{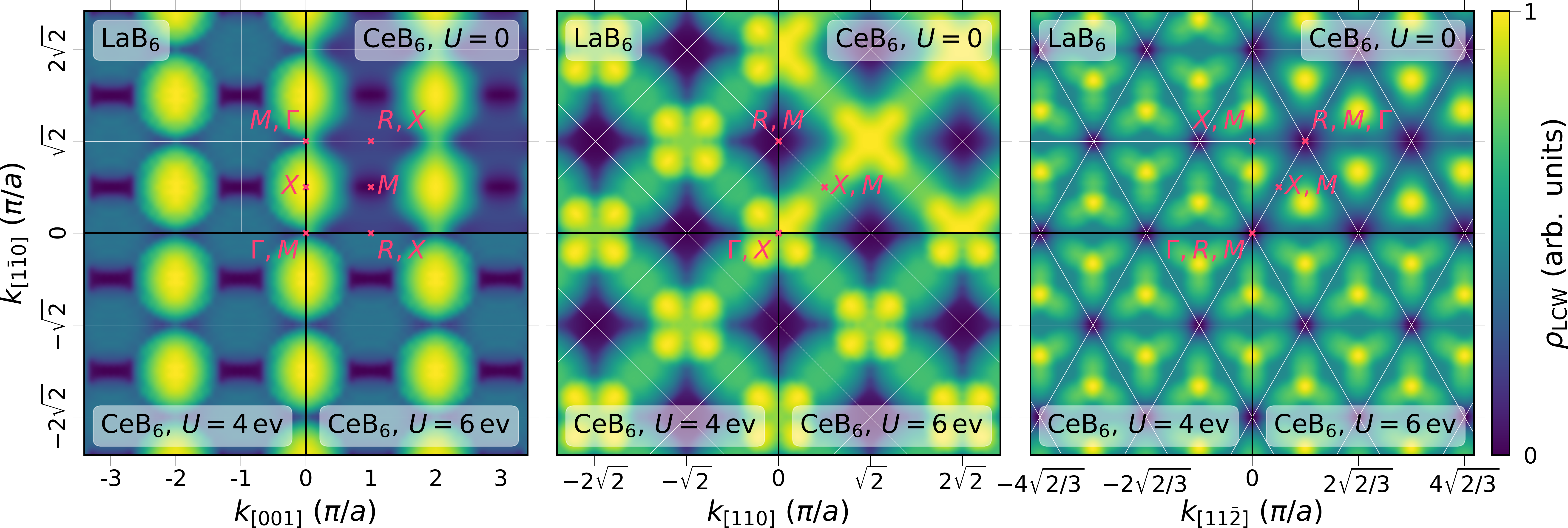}
    \caption{Unconvolved LCW folded back $k$-space density of the calculated ACAR spectra for $\mathrm{LaB}_6$ (upper left) and $\mathrm{CeB}_6$ for ${U=0}$ (upper right), ${U=4}$~eV (bottom left), and ${U=6}$~eV (bottom right). Results correspond to LDA ($\mathrm{LaB}_6$ and $\mathrm{CeB}_6$ ${U=0}$) and LDA+$U$ (the rest) calculations with the Drummond parametrization of the electron-positron enhancement factor.
    The borders of the projected first BZ (white lines) in the repeated zone scheme are shown in all plots.
    From left to right $(110)$, $(001)$, and $(111)$ high symmetry projections.}
    \label{fig:lcw-CeB6-theo}
\end{figure}

In \cref{fig:RA-CeB6-theo} we present results for the calculated radial anisotropy of the 2D-ACAR spectra compared with $\mathrm{CeB}_6$.
The electronic calculations for $\mathrm{LaB}_6$ were done using DFT only.
For $\mathrm{CeB}_6$ we used LDA+$U$ by varying the value of $U$ between $0$ and $6$~eV.
We also show the corresponding LCW back-folded results in \cref{fig:lcw-CeB6-theo}. 
One can clearly see the similarity between the $\mathrm{LaB}_6$ and $\mathrm{CeB}_6$, $U=4$ and $6$~eV spectra.
This can be attributed to the on-site Coulomb repulsion (Hubbard $U$), which pushes unoccupied $4f$ -manifolds above the Fermi level and localizes remained filled $4f$-orbitals.
The localization of a single electron in the $4f$-orbital causes the fully ferromagnetic ground state with the magnetic moment of about $1\mu_B$.

The shown spectra agree with the 2D-ACAR-measurements presented in ref~\cite{bi.fr.97} for both structures ($\mathrm{LaB}_6$ and $\mathrm{CeB}_6$).
We observe that the $\mathrm{CeB}_6$ spectra becomes similar to $\mathrm{LaB}_6$ for $U=4$ and $6$~eV in all shown high-symmetry projections.
This effect is connected to the localization of the $4f$-orbital as its position in energy is shifted further below the Fermi surface with increasing values of the Hubbard $U$ parameter.
On the central panel of the \cref{fig:lcw-CeB6-theo} ($(001)$ projection), where we show the LCW back-folded unconvolved data, one can clearly identify the Fermi surface ``ellipsoids'' in ${[100]}$ and ${[010]}$ directions as well as rounded-square cross-sections of the same ellipsoids along the ${[001]}$ direction.
This demonstrates that the high-resolution 2D-ACAR spectra can be used to deduce bulk FS features of the $R\mathrm{B}_6$ family of compounds, which also including $4f$-valence electron bands.

\section{Summary and Outlook}
\label{sec:summary}

The 2D-ACAR experiments offer answers to some fundamental questions whether the electron momentum 
density in a specific material possesses Fermi surface breaks. 
Through this technique one may provide guidance in searching for a suitable theoretical description of the electronic structure, and thus help in developing an understanding of the mechanism responsible for the occurrence of specific effects, among them also topological features.
Alternatively, the accuracy of any specific electronic structure model can in principle be assessed by comparing the computed 2D-ACAR spectra with the corresponding measurements. 
On the theoretical side, such comparisons have been already made using the most advanced LDA+DMFT method for simple transition metals~\cite{Ceeh2013}, however its extension to correlated multi-orbital $4f$-systems with strong spin-orbit coupling is not currently available.

In the present study we performed 2D-ACAR measurements on single crystalline $\mathrm{LaB}_6$ along three high-symmetry directions. The experimental spectra in both $p$- and $k$-space were compared to LDA calculations and showed good agreement.
In particular the quantitative results extracted from the experiment agree excellently with the corresponding theoretical values of the FS along ${X-M}$, and ${X-R}$ directions; the higher experimental value found for ${X-\Gamma}$, however, is attributed to the fact that this direction is experimentally not directly accessible.

For another member of the rare-earth hexaboride class, the $\mathrm{CeB}_6$ compound with $4f$-states in the valence band, in absence of the recent experimental measurements, the LDA+$U$ method has been applied with different values of the Hubbard parameter $U$. 
The present study was motivated following the proposal that $\mathrm{SmB}_6$, another member of the hexaboride family, may be an interaction driven topological insulator~\cite{dz.su.10,dz.xi.16}. 
The evidence for a metallic surface state in this material is largely accepted, and in the same time low temperature transport~\cite{al.ba.79,zh.bu.13} and (spin-polarized) angle-resolved photo-emission spectroscopy measurements~\cite{dz.xi.16} point towards a bulk small-gap insulating state. Quantum oscillation measurements~\cite{ta.hs.15} found dHvA frequencies, cyclotron masses, and amplitudes that are quite similar to other metallic hexaborides such as $\mathrm{LaB}_6$ and  $\mathrm{CeB}_6$~\cite{is.ta.77,ha.sa.88,on.ko.89}. 
These two compounds exhibit a metallic ground state, involving predominantly conduction electrons with low residual resistivity, and are characterized according to the DFT~(LDA) calculations by a multiply connected Fermi surface of distorted ellipsoids. Band structure calculations for these metallic compounds showed that this type of ``ellipsoids'' are universal Fermi surface features for these materials. In addition, the same type of calculations revealed similar features for $\mathrm{SmB}_6$, when the Fermi level is shifted by hand from the computed position, which is in the gap, either up in the conduction bands or down in the valence bands~\cite{ta.hs.15}.

Based on the presented results and the combined theoretical and experimental (for $\mathrm{LaB}_6$) analysis, we conclude that the ``ellipsoid'' cross-sections ($\alpha$-orbit) and neck sizes increase, hence $\gamma$- and $\varepsilon$-hole orbit of $\mathrm{LaB}_6$ and $\mathrm{CeB}_6$ are significantly reduced in size in comparison with $\mathrm{SmB}_6$.
This is in agreement with the quantum oscillation dHvA measurements, and support the idea that these two compounds are topologically trivial but correlated metals.
Therefore, as a complementary approach to the present LDA(+$U$) treatment of $\mathrm{LaB}_6$ and $\mathrm{CeB}_6$, we might consider the dynamical mean field theory which is able to take into account the full local correlation effects and its extensions also including inter-site correlations. 
Here the role of the crystal electric-field becomes apparent. The sixfold degenerate ${j=5/2}$ multiplet state of $\mathrm{Ce}$-$4f^1$ is split into a $\Gamma_8$ quartet and the $\Gamma_7$ doublet. 
An effective Hamiltonian formulation restricted to the $\Gamma_8$ quartet only, represents a significant reduction of the local Hilbert space, and therefore makes the many-body DMFT computation feasible. 
This will finally allow the description of a truly paramagnetic metallic ground state and may reveal more subtle effects connected to the shape of the Fermi surface.   



\medskip

\textbf{Acknowledgements} \par 
Financial support by the Deutsche Forschungsgemeinschaft through TRR80 (project E2)
Project number 107745057 is gratefully acknowledged.
We acknowledge the fruitful long-term collaboration with Stephen B. Dugdale.
We are very grateful to N. Shitsevalova, from Frantsevich Institute for Problems of Materials Science, National Academy of Sciences of Ukraine, for providing the  $\mathrm{LaB}_6$ single crystal.

\medskip

%
\bibliographystyle{MSP}
\bibliography{bibtex.bib}

\begin{thebibliography}{10}
\providecommand{\url}[1]{\texttt{#1}}
\providecommand{\urlprefix}{URL }

\bibitem{de.fe.97}
L.~Degiorgi, E.~Felder, H.~R. Ott, J.~L. Sarrao, Z.~Fisk,
\newblock \emph{Phys. Rev. Lett.} \textbf{1997}, \emph{79} 5134.

\bibitem{rise.00}
P.~S. Riseborough,
\newblock \emph{Advances in Physics} \textbf{2000}, \emph{49}, 3 257.

\bibitem{an.ar.97}
V.~I. Anisimov, F.~Aryasetiawan, A.~I. Lichtenstein,
\newblock \emph{Journal of Physics: Condensed Matter} \textbf{1997}, \emph{9},
  4 767.

\bibitem{ar.gu.98}
F.~Aryasetiawan, O.~Gunnarsson,
\newblock \emph{Reports on Progress in Physics} \textbf{1998}, \emph{61}, 3
  237.

\bibitem{ko.vo.04}
G.~Kotliar, D.~Vollhardt,
\newblock \emph{Physics Today} \textbf{2004}, \emph{57}, 3 53.

\bibitem{ko.sa.06}
G.~Kotliar, S.~Y. Savrasov, K.~Haule, V.~S. Oudovenko, O.~Parcollet, C.~A.
  Marianetti,
\newblock \emph{Rev. Mod. Phys.} \textbf{2006}, \emph{78}, 3 865.

\bibitem{dz.su.10}
M.~Dzero, K.~Sun, V.~Galitski, P.~Coleman,
\newblock \emph{Phys. Rev. Lett.} \textbf{2010}, \emph{104} 106408.

\bibitem{ch.ta.15}
T.-R. Chang, T.~Das, P.-J. Chen, M.~Neupane, S.-Y. Xu, M.~Z. Hasan, H.~Lin,
  H.-T. Jeng, A.~Bansil,
\newblock \emph{Phys. Rev. B} \textbf{2015}, \emph{91} 155151.

\bibitem{sh.ho.20}
S.-H. Hung, H.-T. Jeng,
\newblock \emph{Materials} \textbf{2020}, \emph{13}, 19.

\bibitem{ho.ko.64}
P.~Hohenberg, W.~Kohn,
\newblock \emph{Phys. Rev.} \textbf{1964}, \emph{136} B864.

\bibitem{kohn.99}
W.~Kohn,
\newblock \emph{Rev. Mod. Phys.} \textbf{1999}, \emph{71} 1253.

\bibitem{jo.gu.89}
R.~O. Jones, O.~Gunnarsson,
\newblock \emph{Rev. Mod. Phys.} \textbf{1989}, \emph{61} 689.

\bibitem{jone.15}
R.~O. Jones,
\newblock \emph{Rev. Mod. Phys.} \textbf{2015}, \emph{87} 897.

\bibitem{Tanaka2002}
K.~Tanaka, Y.~{\={O}}nuki,
\newblock \emph{Acta Crystallographica Section B} \textbf{2002}, \emph{58}, 3
  Part 2 423.

\bibitem{CHEN2004}
C.-H. Chen, T.~Aizawa, N.~Iyi, A.~Sato, S.~Otani,
\newblock \emph{Journal of Alloys and Compounds} \textbf{2004}, \emph{366}, 1
  L6.

\bibitem{jo.ru.87}
W.~Joss, J.~M. van Ruitenbeek, G.~W. Crabtree, J.~L. Tholence, A.~P.~J. van
  Deursen, Z.~Fisk,
\newblock \emph{Phys. Rev. Lett.} \textbf{1987}, \emph{59} 1609.

\bibitem{ne.al.15}
M.~Neupane, N.~Alidoust, I.~Belopolski, G.~Bian, S.-Y. Xu, D.-J. Kim, P.~P.
  Shibayev, D.~S. Sanchez, H.~Zheng, T.-R. Chang, H.-T. Jeng, P.~S.
  Riseborough, H.~Lin, A.~Bansil, T.~Durakiewicz, Z.~Fisk, M.~Z. Hasan,
\newblock \emph{Phys. Rev. B} \textbf{2015}, \emph{92} 104420.

\bibitem{bi.al.96}
M.~Biasini, M.~Alam, S.~Dugdale, H.~Fretwell, Y.~Kubo, H.~Harima, N.~Sato,
\newblock \emph{Applied Surface Science} \textbf{1997}, \emph{116} 335,
  proceedings of the Seventh International Workshop on Slow-Positron Beam
  Techniques for Solids and Surfaces.

\bibitem{bi.fr.97}
M.~Biasini, H.~M. Fretwell, S.~B. Dugdale, M.~A. Alam, Y.~Kubo, H.~Harima,
  N.~Sato,
\newblock \emph{Phys. Rev. B} \textbf{1997}, \emph{56} 10192.

\bibitem{bi.mo.01}
M.~Biasini, M.~Monge, G.~Kontrym-Sznajd, M.~Gemmi, N.~Sato,
\newblock In \emph{Positron Annihilation - ICPA-12}, volume 363 of
  \emph{Materials Science Forum}. Trans Tech Publications Ltd, \textbf{2001}
  582--584.

\bibitem{Weber2015}
J.~A. Weber, A.~Bauer, P.~B\"oni, H.~Ceeh, S.~B. Dugdale, D.~Ernsting,
  W.~Kreuzpaintner, M.~Leitner, C.~Pfleiderer, C.~Hugenschmidt,
\newblock \emph{Phys. Rev. Lett.} \textbf{2015}, \emph{115} 206404.

\bibitem{Ceeh2016}
H.~A. Ceeh, J.~A. Weber, P.~B\"oni, M.~Leitner, D.~Benea, L.~Chioncel,
  H.~Ebert, J.~Minár, D.~Vollhardt, C.~Hugenschmidt,
\newblock \emph{Scientific Reports} \textbf{2016}, \emph{6} 20898.

\bibitem{Jarlborg1987}
T.~Jarlborg, A.~K. Singh,
\newblock \emph{Physical Review B} \textbf{1987}, \emph{36} 4660.

\bibitem{Lock1973}
D.~G. Lock, V.~H.~C. Crisp, R.~N. West,
\newblock \emph{Journal of Physics F: Metal Physics} \textbf{1973}, \emph{3}, 3
  561.

\bibitem{Leitner2012}
M.~Leitner, H.~Ceeh, J.-A. Weber,
\newblock \emph{New Journal of Physics} \textbf{2012}, \emph{14}, 12 123014.

\bibitem{Ceeh2013}
H.~Ceeh, J.~A. Weber, M.~Leitner, P.~Böni, C.~Hugenschmidt,
\newblock \emph{Rev. Sci. Instrum.} \textbf{2013}, \emph{84}, 4 043905.

\bibitem{elk}
J.~K. Dewhurst, S.~Sharma, L.~Nordstr\"{o}m, F.~Cricchio, O.~Granas, E.~K.~U.
  Gross,
\newblock {ELK} - electronic structure code - version 5.2.14,
\newblock \urlprefix\url{http://elk.sourceforge.net}.

\bibitem{pe.wa.92}
J.~P. Perdew, Y.~Wang,
\newblock \emph{Phys. Rev. B} \textbf{1992}, \emph{45} 13244.

\bibitem{pe.ma.03}
A.~G. Petukhov, I.~I. Mazin, L.~Chioncel, A.~I. Lichtenstein,
\newblock \emph{Phys. Rev. B} \textbf{2003}, \emph{67}, 15 153106.

\bibitem{bo.ni.86}
E.~Boro\ifmmode~\acute{n}\else \'{n}\fi{}ski, R.~M. Nieminen,
\newblock \emph{Phys. Rev. B} \textbf{1986}, \emph{34}, 6 3820.

\bibitem{pu.ni.94}
M.~J. Puska, R.~M. Nieminen,
\newblock \emph{Rev. Mod. Phys.} \textbf{1994}, \emph{66} 841.

\bibitem{er.bi.14}
D.~Ernsting, D.~Billington, T.~Haynes, T.~Millichamp, J.~Taylor, J.~Duffy,
  S.~Giblin, J.~Dewhurst, S.~Dugdale,
\newblock \emph{Journal of Physics: Condensed Matter} \textbf{2014}, \emph{26},
  49 495501.

\bibitem{la.ha.10}
J.~Laverock, T.~D. Haynes, M.~A. Alam, S.~B. Dugdale,
\newblock \emph{Phys. Rev. B} \textbf{2010}, \emph{82} 125127.

\bibitem{dr.lo.10}
N.~D. Drummond, P.~L\'opez~R\'{\i}os, C.~J. Pickard, R.~J. Needs,
\newblock \emph{Phys. Rev. B} \textbf{2010}, \emph{82} 035107.

\bibitem{dr.lo.11}
N.~D. Drummond, P.~L\'opez~R\'{\i}os, R.~J. Needs, C.~J. Pickard,
\newblock \emph{Phys. Rev. Lett.} \textbf{2011}, \emph{107} 207402.

\bibitem{ta.hs.15}
B.~S. Tan, Y.-T. Hsu, B.~Zeng, M.~C. Hatnean, N.~Harrison, Z.~Zhu,
  M.~Hartstein, M.~Kiourlappou, A.~Srivastava, M.~D. Johannes, T.~P. Murphy,
  J.-H. Park, L.~Balicas, G.~G. Lonzarich, G.~Balakrishnan, S.~E. Sebastian,
\newblock \emph{Science} \textbf{2015}, \emph{349}, 6245 287.

\bibitem{le.we.16}
M.~Leitner, J.~A. Weber, H.~Ceeh,
\newblock \emph{New Journal of Physics} \textbf{2016}, \emph{18}, 6 063033.

\bibitem{pe.zu.81}
J.~P. Perdew, A.~Zunger,
\newblock \emph{Phys. Rev. B} \textbf{1981}, \emph{23} 5048.

\bibitem{cz.sa.94}
M.~T. Czyzyk, G.~A. Sawatzky,
\newblock \emph{Phys. Rev. B} \textbf{1994}, \emph{49}, 20 14211.

\bibitem{ko.he.16}
A.~Koitzsch, N.~Heming, M.~Knupfer, B.~B{\"u}chner, P.~Y. Portnichenko, A.~V.
  Dukhnenko, N.~Y. Shitsevalova, V.~B. Filipov, L.~L. Lev, V.~N. Strocov,
  J.~Ollivier, D.~S. Inosov,
\newblock \emph{Nature Communications} \textbf{2016}, \emph{7}, 1 10876.

\bibitem{dz.xi.16}
M.~Dzero, J.~Xia, V.~Galitski, P.~Coleman,
\newblock \emph{Annual Review of Condensed Matter Physics} \textbf{2016},
  \emph{7}, 1 249.

\bibitem{al.ba.79}
J.~W. Allen, B.~Batlogg, P.~Wachter,
\newblock \emph{Phys. Rev. B} \textbf{1979}, \emph{20} 4807.

\bibitem{zh.bu.13}
X.~Zhang, N.~P. Butch, P.~Syers, S.~Ziemak, R.~L. Greene, J.~Paglione,
\newblock \emph{Phys. Rev. X} \textbf{2013}, \emph{3} 011011.

\bibitem{is.ta.77}
Y.~Ishizawa, T.~Tanaka, E.~Bannai, S.~Kawai,
\newblock \emph{Journal of the Physical Society of Japan} \textbf{1977},
  \emph{42}, 1 112.

\bibitem{ha.sa.88}
H.~Harima, O.~Sakai, T.~Kasuya, A.~Yanase,
\newblock \emph{Solid State Communications} \textbf{1988}, \emph{66}, 6 603.

\bibitem{on.ko.89}
Y.~\={O}nuki, T.~Komatsubara, P.~H.~P.~Reinders, M.~Springford,
\newblock \emph{Journal of the Physical Society of Japan} \textbf{1989},
  \emph{58}, 10 3698.

\end{thebibliography}










\end{document}